\documentclass[aps,prb,twocolumn,groupedaddress,floatfix]{revtex4}
\usepackage{graphicx}
\usepackage{epsfig}
\begin{document}

\title{Superiority of the $h$-index over the Impact Factor for
Physics}
\author{Casey~W.~Miller}
\affiliation{Department of Physics, University of
California,~San~Diego\\9500 Gilman Drive, La Jolla, CA 92093, USA}

\begin{abstract}
\noindent Focusing specifically on physics periodicals, I show
that the journal Impact Factor is not correlated with Hirsch's
$h$-index.  This implies that the Impact Factor is not a good
measure of research quality or influence because the $h$-index is
a reflection of peer review, and thus a strong indicator of
research quality. The impact gap between multidisciplinary
journals and physics-only journals is significantly reduced when
$h$ is used instead of the Impact Factor.  Additionally, the
impact of journals specializing in review articles is inherently
deflated using $h$ because of the limited number of annual
publications in such periodicals. Finally, a reordering of the top
ranking journals occurs with $h$ when only the physics articles of
multidisciplinary journals are considered, falling more in line
with the average physicist's interpretation of a journal's
prestige.
\end{abstract}

\maketitle

\begin{widetext}\begin{figure}
\begin{center}
\epsfig{file=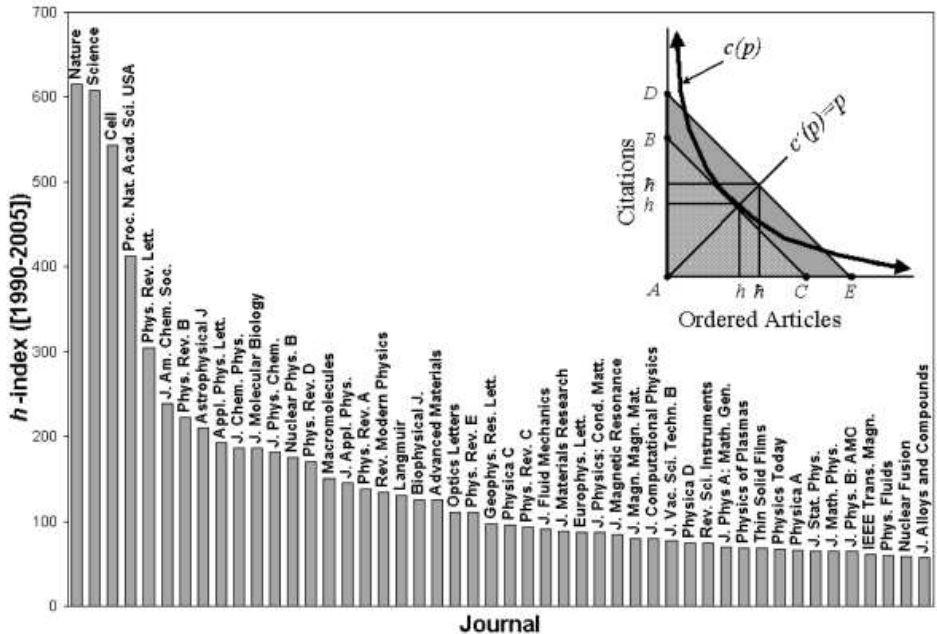,width=5in} %
\caption{\small{(Color online)
     $h$-index for fifty journals from major subfields of physics.
     (Inset) Graphical representation of the indices $h$ and $\hbar$ as defined in the text.}}
    \label{all}
\end{center}
\end{figure}\end{widetext}

The journal Impact Factor (\textit{IF}) has earned the adjective
\textit{notorious} for several reasons \cite{Devouring,
ChronColloq, ChronComp}. First, in contradiction to healthy
philosophies of science, the data used to determine the
\textit{IF} are not publicly available. Second, authors have
reported unethical editorial practices that have the motive of
increasing a journal's \textit{IF}, including editors insisting
that additional references to their journal be added to a
bibliography \textit{before} accepting an article for publication.
Third, publishing in high-\textit{IF} journals is used as a filter
for hiring tenure-track faculty and advancement to tenure.
R.\,Monastersky's ``The number that's devouring science"
highlights other intriguing issues \cite{Devouring}, and astutely
notes that the pragmatic motto ``publish or perish" has mutated to
``publish in a high-impact journal or perish."\\
\indent The impact factor is, in spirit, a reasonable metric: the
total number of citations divided by the total number of articles
\cite{IFcalc}. The three highest \textit{IF} periodicals in which
physicists publish were reported for 2003 as \textit{Nature}
(31.0), \textit{Science} (29.2), and \textit{Reviews of Modern
Physics} (28.2) \cite{JCR}.  Many physicists find it odd that
\textit{Physical Review Letters}, historically the most well
respected physics journal, boasts a meager 7.0. Another indication
that the \textit{IF} rankings are questionable is that experts
disagree with the rankings for their subfields: few of the top
titles are recognized, and even fewer are considered prestigious.
Upon inspection, most of the top \textit{IF} journals in each
subfield specialize in review articles. Review articles are
typically highly cited, though only a few articles are published
annually. This causes such journals' high (and arguably,
anomalous) \textit{IF}, while simultaneously making original
research journals appear relatively weak.\\
\indent Based on the \textit{IF}'s several peculiarities,
alternative measures of journal quality are being sought. In my
opinion, the only reasonable way to rank physics journals is for
physicists to measure the quality of published physics in each
journal. While this may at first seem unrealistic, an indirect
form of this type of peer review already exists in the form of
citations. Articles of high quality or broad interest receive many
more citations than articles of low quality or limited interest,
and are therefore more scientifically influential. Extending this
idea to periodicals, one can inspect the citation and publication
history of a journal in a scientific and unbiased way to determine
the average impact of its publications. I thus propose the
$h$-index as a logical measure of a journal's influence on
science.\\
\indent The $h$-index, developed by J.\,E.\,Hirsch to quantify the
scientific research output of an individual \cite{hfactor}, has
become popular because it is logically sound, simple to
understand, and, most importantly, simple to calculate with easily
obtained data. An individual's $h$-index is determined by
searching a scientific database like Thompson's ISI Web of
Knowledge \cite{ISI} for all articles by an author, and ranking
the output articles by the number of citations such that article 1
has the most citations; $h$ is the rank of the lowest ranking
article whose number of citations is bounded below by its rank.
One can arbitrarily extend this procedure to journals,
departments, institutions, or even zip codes. A graphical
definition of the $h$-index is shown in the inset of
Fig.~\ref{all}. If $c(p)$ is the number of citations for paper $p$
in the ordered list, then $h$ is the intersection of $c(p)$ with
the line $c'(p)=p$.  I have empirically observed that $c(p)$ is
bounded below by a right isosceles triangle with legs of length
$2h$, such as $ABC$; violations are more likely to exist for
scientifically young subjects due to inadequate statistics. An
(approximately) equivalent definition of $h$ is the coordinate of
the intersection of the hypotenuse $BC$ with $c'(p)=p$.\\
\indent The total number of citations can be used to develop a
related, but arguably more poignant index. Consider $ADE$, a
similar triangle of $ABC$, whose area ($\alpha$) equals
$\sum_p\,c(p)$. Using the geometric definition, a new index, which
I cannot resist naming the $\hbar$-index, is the coordinate of the
intersection of the hypotenuse $DE$ with the line $c'(p)=p$, or
mathematically, $\hbar = \sqrt{(\alpha/2)}$.  The $\hbar$-index is
a more comprehensive measure of the overall structure of $c(p)$
for two reasons. First, $\hbar$ incorporates the most highly cited
articles, while $h$ basically ignores all articles with citations
much greater than $h$. Second, $\hbar$ takes into account the body
of articles with moderate numbers of citations, while $h$ again
ignores all such articles.\\
\indent To illustrate the main difference between $h$ and $\hbar$,
consider two similarly aged \cite{age}, fully tenured, condensed
matter physicists from the University of California, San Diego:
L.\,J.\,Sham and I.\,K.\,Schuller.  These professors are both well
respected in this field, and have each received numerous awards
for their scientific achievements. Both have $h=54$, indicating
that they have contributed approximately the same amount to
science. However, $h$ does not take into account that Sham's
article with W. Kohn \cite{KohnSham} that introduced density
functional theory$-$for which Kohn was awarded the Nobel Prize in
1998$-$is the most highly cited paper in the history of
\textit{Physical Review}. Professor Schuller, on the other hand,
has a large body of work and is among the most highly cited
researchers in physics, though his single most highly cited paper
has an order of magnitude fewer citations than Sham's (and neither
he nor his immediate coworkers has yet to win the Nobel Prize). I
find that Sham has $\hbar = 102$, and Schuller has $\hbar = 76$.
Thus the \textit{Sham-Schuller Paradox} is resolved using $\hbar$.
Degeneracies with the $\hbar$-index will of course exist, but
these will not be as qualitatively displeasing as in this and similar cases.\\
\indent For some of the individuals noted in Hirsch's original
article, I find the following for the indices ($h, \hbar$):
E.\,Witten (112, 163), P.\,W.\,Anderson (96, 164), S.\,Weinberg
(89, 139), J.\,N.\,Bahcall (77, 102), D.\,J.\,Scalapino (76, 95),
S.\,G.\,Louie (76, 97), R.\,Jackiw (69, 106), C.\,Vafa (67, 82),
D.\,J.\,Gross (67, 106), and S.\,W.\,Hawking (62, 98).\\
\indent Applying the $h$-index to individuals has proven very
effective. Hirsch unequivocally showed that a large $h$, and more
importantly d$h$/dt, indicates a successful scientist.
Additionally, a recent hind-sight study showed a strong
correlation between $h$ and committee peer review: individuals
that were granted prestigious post-doctoral fellowships
\cite{young} in biomedicine by a committee of well known
scientists in that field from 1990 to 1995 had on average higher
$h$-indices than other applicants \cite{doeshwork}. Upon
reflection, this study indicates the obvious: $h$ measures how
one's contributions are viewed by one's peers. Highly valued
articles will receive many citations from the scientific
community, which will result in a higher $h$-index for the authors
or publishing journal. Similar correlation of $h$ with peer review
was recently demonstrated in chemistry \cite{doeshwork2}.\\
\begin{table}[t]\vspace{-0.1in}
\caption{\small{$h$, $\hbar$, 2003 Impact Factor, total number of
publications, and the percent of uncited articles as of late
March, 2006 for the subset of a journal's publications noted in
the text. $^a$Based on title. $^b$Based on automated filter
\cite{Filter}.}}
\begin{ruledtabular}
 \begin{tabular}{rccccc}
\vspace{-0.05in} \tiny \underline{JOURNAL} &\tiny \underline{
\,\,\,$h$\,\,\,}&\tiny \underline{\,\,\,$\hbar$\,\,\,}&\tiny
\underline{\textit{2003 IF}}&\tiny \underline{TOTAL}&\tiny \underline{UNCITED}\\
\vspace{-0.1in}\tiny Nature  &\tiny616 &\tiny1157    &\tiny31.0&\tiny    39322&\tiny   32\\
\vspace{-0.1in}\tiny Science &\tiny608&\tiny 1110   &\tiny 29.2&\tiny 34361  &\tiny 30\\
\vspace{-0.1in}\tiny Cell  &\tiny  543&\tiny 864 &\tiny26.6 &\tiny   7030&\tiny  5\\
\vspace{-0.1in}\tiny Proc. Natl. Acad. Sci. U.S.A.   &\tiny413 &\tiny1175 &\tiny   10.3  &\tiny  45192  &\tiny 6\\
\vspace{-0.08in}\tiny Phys. Rev. Lett.   &\tiny 305 &\tiny869 &\tiny7.0&\tiny 46461&\tiny 8\\
\vspace{-0.08in}\tiny Nature (only physics)  &\tiny 263$^a$, 285$^b$ &\tiny 499$^b$ &\tiny- &\tiny  - &\tiny -\\
\vspace{-0.1in}\tiny Science (only physics) &\tiny 260$^a$&\tiny - &\tiny  - &\tiny  - &\tiny  - \\
\vspace{-0.1in}\tiny J. Am. Chem. Soc. &\tiny239 &\tiny815&\tiny 6.5 &\tiny39513 &\tiny6 \\
\vspace{-0.1in}\tiny Phys. Rev. B   &\tiny 223&\tiny 796&\tiny 3.0&\tiny 77331  &\tiny 12 \\
\vspace{-0.1in}\tiny Astrophys. J. &\tiny210 &\tiny654&\tiny 6.6 &\tiny35318 &\tiny  6 \\
\vspace{-0.1in}\tiny Appl. Phys. Lett. &\tiny193 &\tiny615 &\tiny4.0 &\tiny44480  &\tiny 13 \\
\vspace{-0.1in}\tiny J. Chem. Phys. &\tiny187 &\tiny623 &\tiny3.0 &\tiny39701   &\tiny9 \\
\vspace{-0.1in}\tiny J. Mol. Biol. &\tiny187 &\tiny- &\tiny5.2 &\tiny12898 &\tiny  7 \\
\vspace{-0.1in}\tiny J. Phys. Chem.  &\tiny181 &\tiny646 &\tiny3.3 &\tiny47702 &\tiny11 \\
\vspace{-0.1in}\tiny Nucl. Phys. B &\tiny175&\tiny 386 &\tiny5.3&\tiny 13888 &\tiny  28 \\
\vspace{-0.1in}\tiny Phys. Rev. D &\tiny170&\tiny 484 &\tiny4.6&\tiny 27902&\tiny   10\\
\vspace{-0.1in}\tiny Macromolecules &\tiny 151&\tiny 475&\tiny 3.6 &\tiny20954&\tiny   7 \\
\vspace{-0.1in}\tiny J. Appl. Phys.&\tiny 146&\tiny 493&\tiny 2.2 &\tiny44925 &\tiny  15 \\
\vspace{-0.1in}\tiny Phys. Rev. A &\tiny138 &\tiny426 &\tiny2.6 &\tiny26202&\tiny 12 \\
\vspace{-0.1in}\tiny Rev. Mod. Phys.&\tiny 134 &\tiny188 &\tiny28.2 &\tiny573 &\tiny6 \\
\vspace{-0.1in}\tiny Langmuir   &\tiny 130&\tiny-&\tiny 3.1 &\tiny18201 &\tiny11 \\
\vspace{-0.1in}\tiny Biophys. J.  &\tiny126 &\tiny333 &\tiny4.5&\tiny 9572 &\tiny9 \\
\vspace{-0.1in}\tiny Adv. Mater. &\tiny125&\tiny - &\tiny7.3&\tiny 4659&\tiny    15 \\
\vspace{-0.1in}\tiny Opt. Lett. &\tiny111 &\tiny280 &\tiny3.4&\tiny 11073 &\tiny13 \\
\vspace{-0.1in}\tiny Phys. Rev. E &\tiny111 &\tiny363 &\tiny2.2&\tiny 26660 &\tiny16 \\
\vspace{-0.1in}\tiny Geophys. Res. Lett.&\tiny 97&\tiny  -&\tiny 2.4 &\tiny16313 &\tiny16 \\
\vspace{-0.1in}\tiny Physica C  &\tiny 96&\tiny - &\tiny1.2 &\tiny13567 &\tiny18 \\
\vspace{-0.1in}\tiny Phys. Rev. C &\tiny93 &\tiny-&\tiny 2.7 &\tiny13407 &\tiny11 \\
\vspace{-0.1in}\tiny J. Fluid Mech.&\tiny 91 &\tiny - &\tiny1.8&\tiny 5958 &\tiny11 \\
\vspace{-0.1in}\tiny J. Mater. Res. &\tiny88&\tiny - &\tiny1.6 &\tiny7141 &\tiny15 \\
\vspace{-0.1in}\tiny Europhys. Lett. &\tiny87 &\tiny- &\tiny2.1 &\tiny7872 &\tiny14 \\
\vspace{-0.1in}\tiny J. Physics: Cond. Mat. &\tiny87&\tiny - &\tiny1.8 &\tiny18579 &\tiny18 \\
\vspace{-0.1in}\tiny J. Magn. Resonance &\tiny85 &\tiny- &\tiny2.1 &\tiny3374 &\tiny12 \\
\vspace{-0.1in}\tiny J. Magn. Magn. Mat.&\tiny 80 &\tiny- &\tiny0.9 &\tiny18138 &\tiny30 \\
\vspace{-0.1in}\tiny J. Comput. Phys. &\tiny79 &\tiny- &\tiny1.8 &\tiny3835 &\tiny16 \\
\vspace{-0.1in}\tiny J. Vac. Sci. Technol. B &\tiny77 &\tiny-&\tiny 1.6 &\tiny10213 &\tiny18 \\
\vspace{-0.1in}\tiny Physica D &\tiny75 &\tiny- &\tiny1.6 &\tiny2673 &\tiny8 \\
\vspace{-0.1in}\tiny Rev. Sci. Instrum. &\tiny75&\tiny - &\tiny1.3 &\tiny13959 &\tiny23 \\
\vspace{-0.1in}\tiny J. Phys A: Math. Gen. &\tiny70&\tiny - &\tiny1.4 &\tiny13144 &\tiny20 \\
\vspace{-0.1in}\tiny Phys. Plasmas&\tiny69 &\tiny- &\tiny2.1 &\tiny7672 &\tiny19 \\
%
\vspace{-0.1in}\tiny Thin Solid Films&\tiny 69 &\tiny- &\tiny1.6&\tiny 17023 &\tiny22 \\
\vspace{-0.1in}\tiny Phys. Today   &\tiny67 &\tiny-&\tiny 5.0&\tiny 5652 &\tiny81 \\
\vspace{-0.1in}\tiny Physica A &\tiny66&\tiny - &\tiny1.2 &\tiny5374 &\tiny15 \\
\vspace{-0.1in}\tiny J. Stat. Phys. &\tiny 65&\tiny - &\tiny1.2 &\tiny3652 &\tiny18 \\
\vspace{-0.1in}\tiny J. Math. Phys. &\tiny65&\tiny - &\tiny1.5 &\tiny6975 &\tiny23 \\
\vspace{-0.1in}\tiny J. Phys. B: AMO &\tiny65 &\tiny -&\tiny 1.7&\tiny 8493 &\tiny12 \\
\vspace{-0.1in}\tiny IEEE Trans. Magn.&\tiny 61&\tiny - &\tiny1.0 &\tiny16283 &\tiny29 \\
\vspace{-0.1in}\tiny Phys. Fluids &\tiny60 &\tiny -&\tiny 1.6 &\tiny4924 &\tiny20 \\
\vspace{-0.1in}\tiny Nucl. Fusion&\tiny 59&\tiny - &\tiny3.4 &\tiny2914 &\tiny15 \\
\vspace{-0.1in}\tiny J. Alloys and Compounds &\tiny57 &\tiny- &\tiny1.1 &\tiny12795 &\tiny25 \\
\vspace{-0.1in}\tiny J. Physics D: Appl. Phys.  &\tiny 55&\tiny - &\tiny1.3&\tiny 7597 &\tiny20\\
\tiny Plasma Phys. Contr. Fusion &\tiny49 &\tiny - &\tiny 2.8
&\tiny2946 &\tiny18
 \end{tabular}
 \end{ruledtabular} \label{table}
 \end{table}
 \indent I compared the rankings of scientific periodicals
frequently targeted by physicists from a variety of subfields
using $h$ and $\hbar$ to evaluate the \textit{IF} as a measure of
research quality. The data used to calculate both indices were
obtained using ISI to search by ``source", limited to 1990-2006,
and the following document types: Article, Letter, Review,
Correction, Editorial Material, or Note. The latter three were
included because comments, errata, and retractions are listed
under these headings. Journals that began publishing after 1990
were excluded \cite{exceptions}. I chose sixteen years as an
estimate of the time between entering graduate school and
receiving tenure at a university. Additionally, this long
timescale reduces the influence of high frequency fluctuations,
such as those due to spectacular claims that are often disproved
soon after their initial publication (the \textit{IF} only uses
two years \cite{IFcalc}). The data were harvested in a 10 day
window in late March, 2006; results are presented in Table~\ref{table}.\\
\indent Figure~\ref{all} shows the $h$-index of fifty
physics-related periodicals. As I discuss below, the relatively
large $h$ of the top four is due to the fact that these
periodicals publish articles from other scientific disciplines,
which inflates their indices relative to physics-only journals.
\textit{Physical Review Letters} also has a jump relative to the
remaining physics journals. This is probably because it publishes
work from all of physics, while
the others are for specific subfields.\\
\begin{figure}
\begin{center}
\epsfig{file=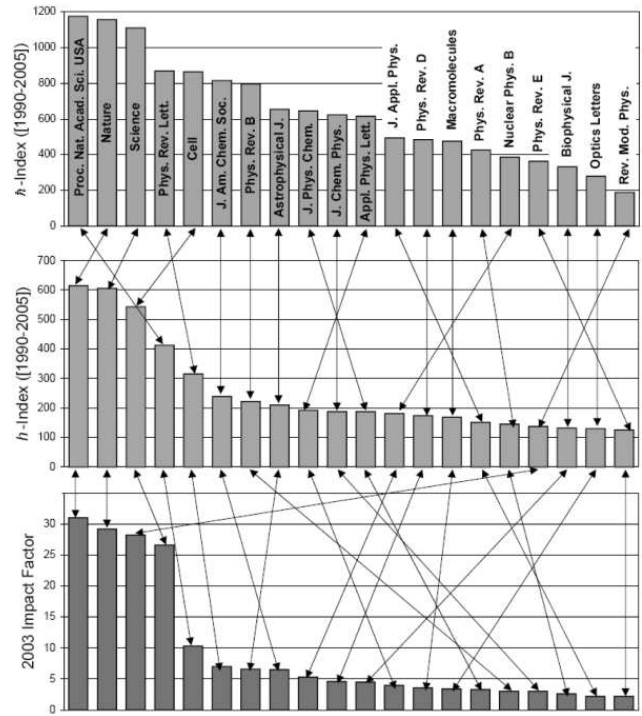,width=3.3in} %
     \caption{\small{(Color online)
     $\hbar$, $h$, and the 2003 \textit{IF} for twenty physics journals
     with $h\,>\,100$. The arrows show the reordering between the indices.}}
    \label{arrows}
\end{center}
\end{figure}
\indent Figure \ref{arrows} shows the journals ranked using
$\hbar$, $h$, and the 2003 \textit{IF} as published in the Journal
Citation Report \cite{JCR} for twenty titles with $h\,>\,100$. The
arrows show how the individual journals change rank when analyzed
with the different indices. A journal's rank changes by an average
of 1.3 positions between $\hbar$ and $h$, 4.4 between $\hbar$ and
\textit{IF}, and 3.4 between $h$ and \textit{IF}. The
\textit{IF}-based rankings are inconsistent with $h$ and $\hbar$.
\textit{This means the \textit{IF} is inconsistent with peer
review, and is direct evidence that the \textit{IF} is not a good
measure of the quality of physics published in a particular
journal.} Interestingly, the greatest leap of $-17$ places was
made by \textit{Reviews of Modern Physics}. This may indicate that
$h$ intrinsically solves the impact inflation problem of review
journals.\\
\indent The large discrepancy between indices for \textit{Nature}
and \textit{Science} relative to physics-only journals is due to
the fact that their indices are inflated by work from other
scientific fields \cite{notCell}.  Evidence of this inflation
includes: 1) I can only identify $\sim8\%$ of the top $h$ articles
in \textit{Nature} and \textit{Science} as physics research based
on a qualitative assessment of their content by their titles
\cite{topics}.  2) Physics articles represent only a small
fraction of the total number of articles in these periodicals,
which means fewer physics articles per unit time. The total number
of articles is important in determining $h$ and $\hbar$.  Fewer
articles means fewer citations, and thus an inherently lower
$h$-index (as seen with \textit{Reviews of Modern Physics}).
Together, these points indicate that the impact of
multidisciplinary journals, regardless of the metric, is heavily
weighted by topics other than physics.\\
\indent An important question naturally presents itself:
\textit{how does the subset of physics articles in
multidisciplinary periodicals compare with physics-only journals?}
To investigate this, an automated filter was used to select
``physics" articles from the aforementioned data set of
\textit{Nature} publications. The selection criteria were based
solely on the references of each article. An article initially
qualified as physics if it contained at least one physics
reference. The percentage of an article's total references that
qualified as physics references served as a final, tunable filter.
A reference was deemed a ``physics reference" if it contained
generic search strings (\verb"astron", \verb"biophys",
\textit{etc.}) or specific abbreviations (\verb"phys fluids",
\verb"j phys b", \textit{etc.}), and did not contain other strings
(\verb"physio", \verb"rehab", \textit{etc.}) \cite{Filter}. The
filter excluded \verb"nature" and \verb"science" because a
disproportionate number of references contain at least one of
these strings, which would have undermined this analysis. This
method is certainly not perfect, but it is robust in the sense
that perturbing the search strings does not significantly alter
the results; the indices change by less than
five percent when \verb"phys rev" is purposefully excluded with the filter.\\
\begin{figure}
\begin{center}
\epsfig{file=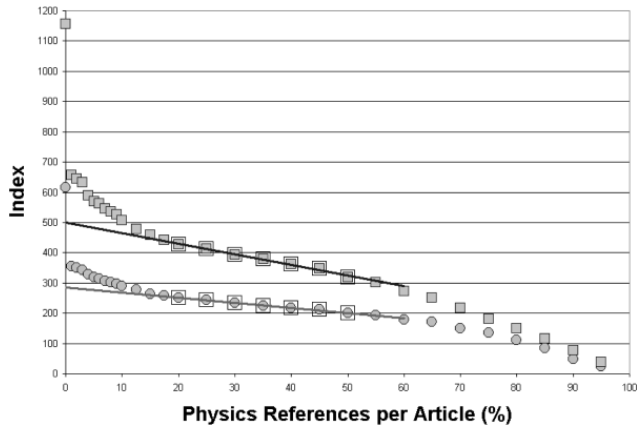,width=3.3in} %
     \caption{\small{(Color online)
     $h$ (lower, circles) and $\hbar$ (upper, squares) indices for \textit{Nature}
     using the percent of physics references in individual publications as a filter.
     The highlighted data were used for the linear fit, which yields
     $h_{phys}\,=\,285$ and $\hbar_{phys}\,=\,499$ when extrapolated to the
     ordinate-intercept.}}
    \label{percent}
\end{center}
\end{figure}
\indent Figure~\ref{percent} shows $h$ and $\hbar$ for the
qualifying \textit{Nature} articles as a function of the final
filter percentage. Each point on these curves was obtained by 1)
selecting all articles whose bibliographies contained \textit{at
least} n\% physics references, then 2) determining $h$ and $\hbar$
for this subset of articles. The tail below fifteen percent is
primarily due to physics references in non-physics articles. Both
indices decay with increasing filter strictness as a result of the
decreasing number of qualifying articles. The data around the
inflection points (20-50\%) were fit to lines, and the
ordinate-intercepts were used to define physics-specific indices.
I thus determined $\hbar_{phys}\,=\,499$, and $h_{phys}\,=\,285$.
The latter is in good agreement with the manual analysis based on
article titles: $h_{phys}\,=\,263$ ($h_{phys}\,=\,260$ for
\textit{Science}). With $\hbar\,=\,869$ and $h\,=\,305$,
\textit{Physical Review Letters} ranks higher than \textit{Nature}
(and presumably \textit{Science}) for both indices. The origin of
the large discrepancy between $\hbar_{phys}$ and $\hbar$ for
\textit{Nature} and \textit{Physical Review Letters},
respectively, is probably
due to the large percentage of uncited articles in \textit{Nature} (see Table~\ref{table}).\\
\indent In summary, the journal \textit{IF} was shown to be
inconsistent with the $h$-index. Insofar as $h$ reflects peer
review, and peer review reflects research quality, these results
indicate that the \textit{IF} is a poor measure of research
quality.  An additional benefit of $h$ is its intrinsic grounding
of the impact of review journals, whose \textit{IF} is anomalously
enhanced by their few annual publications. I showed that the
impact of multidisciplinary journals is enhanced relative to
physics-only journals by the more numerous non-physics articles
contained therein. An analysis of the physics subset of
\textit{Nature} and \textit{Science} revealed that
\textit{Physical Review Letters} has a greater $h$ than either of
these article subsets. This is in accord with \textit{Physical
Review Letters}' historical status as the most reputable physics
journal. Based on these many observations, I conclude that the
indices $h$ and $\hbar$ are superior to the journal \textit{IF} as
indications of the quality of research published in a journal.\\
\begin{acknowledgments}
\vspace{-0.15in} Special thanks to M.\,D.\,Chabot, T.\,Gredig,
Z.-P.\,Li, T.\,C.\,Messina, and W.\,F.\,Egelhoff,\,Jr., for useful
comments. This investigation was performed in the author's spare
time, and was not supported by any funding agency.
\end{acknowledgments}


\begin{thebibliography}{15}
\bibitem[{young()}]{young}
\bibinfo{note}{One must of course cautiously apply $h$, or any such metric, to
young researchers since it will be based on only a few papers, and
is undoubtedly more a reflection of pedigree than potential.}
\bibitem[{age()}]{age}
\bibinfo{note}{This point is disputed by IKS, but it is accurate to $\sim15\%$.}
\end{thebibliography}
\end{document}